\documentclass[twocolumn,showpacs,preprintnumbers,amsmath,amssymb,nofootinbib]{revtex4}

\newcommand{\be}{\begin{equation}}
\newcommand{\ee}{\end{equation}}
\newcommand{\bea}{\begin{eqnarray}}
\newcommand{\eea}{\end{eqnarray}}
\newcommand{\ba}{\begin{array}}
\newcommand{\ea}{\end{array}}

\def\bbox{{\,\lower0.9pt\vbox{\hrule \hbox{\vrule height 0.2 cm
\hskip 0.2 cm \vrule height 0.2 cm}\hrule}\,}}
\newcommand{\dsl}{\pa \kern-0.5em /}

\renewcommand{\d}{\textrm{d}}

\newcommand{\e}{\textrm{e}}
\renewcommand{\d}{\textrm{d}}

\newcommand{\SU}{\mathop{\rm SU}}
\newcommand{\SO}{\mathop{\rm SO}}

\newcommand{\ISO}{\mathop{\rm ISO}}

\usepackage{graphicx}
\usepackage{dcolumn}
\usepackage{bm}

 \usepackage{latexsym}
 \usepackage{epsf}
 \usepackage{amssymb}

 \usepackage{amsmath}
 \usepackage{slashed,epsfig}
 \usepackage{bbm}
 \usepackage{bbm}
 \usepackage{amsmath,amssymb,amsthm}
 \usepackage{pst-all}
 \usepackage{here}

\begin{document}

\preprint{MAD-TH-2008-13, SU-ITP-08/27, UG-FT-241/08, CAFPE-111/08}

\title{Minimal Simple de Sitter Solutions}

\author{Sheikh Shajidul Haque}
\affiliation{Department of Physics, University of
Wisconsin, Madison, WI 53706, USA}
\author{Gary Shiu}
\affiliation{Department of Physics, University of
Wisconsin, Madison, WI 53706, USA}
\affiliation{Department of Physics and SLAC, Stanford University, Stanford, CA 94305, USA}
\author{Bret Underwood}
\affiliation{Department of Physics, McGill University
Montr\'eal, QC H3A 2T8 Canada}
\author{Thomas Van Riet}\vspace{1.2cm}
\affiliation{Departamento de F\'isica Te\'orica y del Cosmos,
Universidad de Granada, 18071 Granada, Spain}

\affiliation{ Departamento de F\'isica, Universidad de Oviedo, Avda.
Calvo Sotelo 18, 33007 Oviedo, Spain}

\begin{abstract}
We show that the minimal set of necessary ingredients to construct
explicit, four-dimensional de Sitter solutions from IIA string
theory at tree-level are O6-planes, non-zero Romans mass parameter,
form fluxes, and negative internal curvature. To illustrate our
general results, we construct such minimal simple de Sitter
solutions from an orientifold compactification of compact hyperbolic
spaces. In this case there are only two moduli and we demonstrate
that they are stabilized to a sufficiently weakly coupled and large
volume regime. We also discuss generalizations of the scenario to
more general metric flux constructions.


\end{abstract}

\maketitle

\section{Introduction}

Among the most surprising and puzzling discoveries in modern physics
is the apparent acceleration of our current universe
\cite{Perlmutter:1998np,Riess:1998cb}. These striking cosmological
observations, together with the associated conceptual issues in
quantum  gravity, have fueled a decade of studies of de Sitter space
from string theory. By now, different strategies in constructing
metastable de Sitter vacua have been suggested in various string
theory limits \cite{Kachru:2003aw, Maloney:2002rr,
Balasubramanian:2005zx, Saltman:2004sn, Saltman:2004jh,
Saueressig:2005es, Davidse:2005ef, Silverstein:2007ac, Palti:2008mg,
Misra:2007yu}. Such continuing efforts in scenario building also
lend support to the picture of  a string landscape
\cite{Susskind:2003kw,Bousso:2000xa,Feng:2000if} realizing,  in an
interesting microphysical way, Weinberg's earlier insight on the
cosmological constant problem \cite{Weinberg:1987dv}.

In light of the observational evidence for an accelerating universe
and the need for a concrete formulation of a dS/CFT correspondence
\cite{Strominger:2001pn}, it is of pragmatic importance to construct
{\it explicit} de Sitter solutions from string theory. Attempts to
construct fully explicit examples are severely hampered by a myriad
of moduli. For example, in many constructions such as
\cite{Kachru:2003aw}, non-perturbative effects which are difficult
to explicitly compute are often invoked to stabilize moduli.
Although in other setups, purely ``perturbative" ingredients, e.g.,
fluxes, are shown to be sufficient in stabilizing all geometric
moduli \cite{DeWolfe:2005uu}, such ingredients only lead us  to an
anti de Sitter minimum. Additional supersymmetry breaking localized
sources, such as anti D-branes \cite{Kachru:2003aw}, KK5-branes
and/or NS5-branes \cite{Hertzberg:2007wc, Silverstein:2007ac} are
then introduced to uplift the vacuum energy \footnote{One can
consider, instead of branes, uplifting by 4D effective field theory
ingredients such as D-term \cite{Burgess:2003ic}. However, such
uplifting D-term vanishes in the absence of an F-term
\cite{Villadoro:2005yq} so one still needs additional ingredients
for completeness of the model.}. Other than backreaction of these
extended objects, the explicit supersymmetry breaking sources
introduced by such uplifting branes make it hard to analyze and
control the corrections to the moduli stabilizing potential. It thus
remains a major challenge to construct fully explicit and
controllable de Sitter vacua  from string theory, especially ones
which admit not only a 4D effective field theory description but can
be analyzed at the level of 10D supergravity equations of motion.

Recently, an interesting scenario was suggested in
\cite{Silverstein:2007ac} by cleverly combining the virtues of
\cite{Maloney:2002rr,Saltman:2004sn} and of \cite{DeWolfe:2005uu}.
The strategy of using manifolds with negative scalar curvature to
generate a positive energy density explored in
\cite{Maloney:2002rr,Saltman:2004sn} was maintained, but with the
internal space replaced by a simpler compactification (more
precisely, a particular twisted torus being a product of two Nil
3-manifolds \cite{Silverstein:2007ac}), such that the machineries
developed for toroidal orientifolds in \cite{DeWolfe:2005uu} can be
easily generalized and applied. It was further argued that with
additional ingredients including KK5-branes and discrete Wilson
lines which are supported on such internal spaces, a metastable de
Sitter vacuum with small cosmological constant can be obtained.
Furthermore, this construction has already motivated new
inflationary  scenarios which can give rise to detectable tensor
modes \cite{Silverstein:2008sg}. In view of these potential
applications, it is of interest to understand what really makes such
constructions tick.

In this paper, we determine what is the minimal set of ingredients
that is truly necessary for the construction of metastable de Sitter
vacua. Upon closer investigation, we found simpler ways to construct
such solutions without invoking the aforementioned localized
uplifting sources and discrete Wilson lines. For simplicity, we use
hyperbolic spaces as an example to illustrate that such {\it minimal
simple de Sitter solutions} exist. Hyperbolic spaces admit no
deformations other than an overall rescaling of their sizes, making
it easier to search for vacua. In principle, one can extend our
analysis to other negatively Ricci curved compactifications which
admit deformations other than the omnipresent dilaton and volume
moduli. In fact, such extension may be useful for constructing de
Sitter vacua whose energy scale of supersymmetry breaking is
parametrically below the compactification scale \footnote{For
hyperbolic spaces which  are rigid, supersymmetry is broken at the
compactifcation scale.}. However, these additional moduli typically
lead to new runaway directions unless the scalar potential  has the
right moduli dependence whose criteria we will briefly sketch below.
We leave the search for such examples for future work.

Various no-go theorems exist for the construction of stable de
Sitter vacua \cite{Hertzberg:2007wc}, based on the consideration of
two-dimensional slices in the full moduli space parametrized by the
volume and the dilaton moduli of the compactification \footnote{See
also \cite{Covi:2008ea} for more advanced no-go theorems in the
effective field theory approach.}. By revisiting the assumptions
made in the arguments, we find that the minimal set of ingredients
needed to add to the usual Ramond (RR) and Neveu-Schwarz (NSNS) fluxes 
and O6/D6 sources to
construct metastable dS vacua in Type IIA string theory are
geometric fluxes and a non-vanishing Romans parameter.

The simplicity of our constructions also motivates us to go beyond
the schematic mechanisms demonstrated in \cite{Silverstein:2007ac},
and to carry out fully explicit computations for complete models.
When the full moduli dependence is restored, our explicit systematic
analysis also make it clear under what conditions can all moduli be
stabilized with the minimal ingredients we identified, and the role
each additional ingredient (KK5-branes, discrete Wilson lines, etc)
play in forbidding new runaway directions.

The minimalism of our de Sitter solutions has several advantages.
First of all, the fact that only three-level ingredients are invoked
makes it easier to calculate from first principle the moduli
stabilizing potential, including numerical factors. As we will also
demonstrate, the moduli are stabilized to a sufficiently large
volume, weak coupling regime so it is self consistent to ignore
higher corrections. Secondly, the absence of uplifting branes frees
us from the concern of their backreaction which is notoriously
difficult to compute. The only backreaction in our model is coming
from the O6 plane. The back reaction of the fluxes is incorporated
in the 4D effective theory in the reductions we consider here.
Finally, the simplicity and explicitness of our solutions enable us
to analyze the system directly from the perspective of the 10D
equations of motion without the crutch of 4D effective field
theories.

\section{Scalar Potential by Dimensional Reduction}
\label{scalarpotential}

We will dimensionally reduce massive IIA supergravity, with the
action in string frame given by (in the conventions of
\cite{Silverstein:2007ac})
\begin{eqnarray}
\label{eq:IIAAction}S&& = \tfrac{1}{2\kappa_{10}^2}\int
\e^{-2\phi}\Bigl(\star \mathcal{R} + 4 \star\d\phi\wedge\d\phi
-\tfrac{1}{2}\star H_3\wedge H_3\Bigr)\nonumber\\&& -\star F_2\wedge
F_2-\star F_4\wedge F_4 -\star m^2+\mbox{CS}+\mbox{(sources)}
\end{eqnarray}
where $2\kappa^2_{10}=(2\pi)^7(\alpha')^4$ and the field strengths
are defined as
\begin{align}
H_3&=\d B_2\,,\\
F_2&=\d C_1+mB_2\,,\\
F_4&=\d C_3 - C_1\wedge H_3-\frac{m}{2}B\wedge B\,,
\end{align}
and the Chern-Simons term (CS) reads
\begin{equation}- \d C_3\wedge \d C_3\wedge B_2+\frac{m}{3}B\wedge B\wedge
B\wedge \d C_3 -\frac{m^2}{20}B\wedge B\wedge B\wedge B\wedge B\,.
\end{equation}

In this paper we are working in the supergravity (plus localized
sources such as D-branes and O-planes) limit; for explicit
solutions, we can and will explicitly check whether this assumption
is valid, finding that we are indeed in the large volume and
(marginally) small string coupling limit, justifying our usage of
the tree level IIA supergravity action and its dimensionally reduced
effective potential (1).  We will  be calling this the tree-level
limit, in contrast to other moduli stabilization techniques such as
KKLT \cite{Kachru:2003aw} which require explicit 4-dimensional
non-perturbative effects.

The dimensional reduction of the
action (\ref{eq:IIAAction}) leads to several terms which
contribute to the effective 4-dimensional scalar potential.
\begin{align}
V &= V_{\text{metric}} + V_3^{NS} + \sum_p V_p^{RR}\nonumber\\ &  +
V_{O6} + V_{D6} + V_{NS5} + V_{KK5}\, .
\label{eq:PotentialSchematic}
\end{align}
where we schematically denoted the contributions coming from the
metric flux, the B-field flux, the RR fluxes, space-filling O6
planes, D6 branes, KK5 and NS5 branes. Before delving into the
details of these contributions to the potential, let us consider
some general properties of the potential
(\ref{eq:PotentialSchematic}) and its capacity for de Sitter vacua.

We will be largely interested in two (real, dimensionless) moduli,
the volume modulus $\rho$ and the dilaton $\tau$ defined by
\begin{equation}
\rho \equiv (\mbox{Vol})^{1/3}\,, \, \, \tau \equiv e^{-\phi}
(\mbox{Vol})^{1/2}\, .
\end{equation}
Additional moduli exist, but depend on the specific model under
consideration.

Recently, it was shown that a ``no-go theorem" exists for inflation and
de Sitter vacua in IIA string theory \cite{Hertzberg:2007wc}.  The
no-go theorem proves that for a {\it vanilla} subset of the possible
contributions to the scalar potential in
(\ref{eq:PotentialSchematic}) (namely NSNS fluxes, RR fluxes, and
O6/D6 sources) there is a bound on the derivatives of the potential,
\begin{equation}
-\rho \frac{\partial V}{\partial\rho}-3\tau \frac{\partial
V}{\partial \tau} = 9 V + \sum_p p V_p \geq 9 V\, . \label{eq:NoGo}
\end{equation}
For inflation, this leads to a bound on the slow roll parameter
$\epsilon \geq \frac{27}{13}$ whenever $V > 0$.  For vacua, we find
from (\ref{eq:NoGo}) that $V = -(\sum_p p V_p)/9$.  As long as
$V_p > 0$, we cannot obtain de Sitter vacua in these models.

A simple way to check for deSitter vacua in IIA models by simple
inspection of the scalar potential is to arrange the contributions
to the potential in the following way,
\begin{equation}\label{eq:potentialterms}
V= a(\rho,M) \tau^{-2} - b(\rho,M) \tau^{-3} + c(\rho,M)
\tau^{-4}\,.
\end{equation}
The quantity $a(\rho,M)$ contains contributions from the curvature of the
internal manifold,
NS 3-form flux, NS 5-branes and KK 5-branes ,
\begin{equation}
a(\rho,M) = \frac{\tilde{C}_f(M)}{\rho}  +
\frac{\tilde{A}_{KK5}(M)}{\rho} +  \frac{\tilde{A}_{NS5}(M)}{\rho^2}
+ \frac{\tilde{A}_{H3}(M)}{\rho^3} \, .
\end{equation}
The quantity $b(\rho,M)$ contains contributions from O6-planes and
D6-branes,
\begin{equation}
b(\rho,M) = +n_{O6}f(M)- n_{D6} g(M) \, ,
\end{equation}
where $f$ and $g$ are functions of moduli different from $\rho$ and
$\tau$. The quantity $c(\rho,M)$ contains contributions from the RR
fluxes (and by extension, fractional Wilson lines),
\begin{equation}
c(\rho,M) = \rho^3 \tilde{m}^2 + \rho \tilde{A}_2(M) +
\frac{\tilde{A}_4^{elec}(M)}{\rho} + \frac{\tilde{A}_6(M)}{\rho^3}\, .
\end{equation}

As described in \cite{Silverstein:2007ac} and discussed in more detail in
appendix \ref{sec:appendix}, finding de Sitter vacua
with small cosmological constant (cc) is as easy as using the
``a,b,c" quantities to search for critical points of
\begin{equation}
\frac{4 a c}{b^2} \approx 1\, . \label{eq:EvaCritical}
\end{equation}
In particular, writing
\begin{equation}
\frac{4 a c}{b^2} = 1 + \delta(\rho,M)
\end{equation}
and denoting minimized quantities with a subscript $_0$, for
$\delta_0 \approx 0$ we have a vacuum solution with positive vacuum
energy
\begin{equation}
V_{min} \approx \left(\frac{b_0}{2 c_0}\right)^4 c_0 \delta_0\, .
\end{equation}

It is straightforward to show the no-go theorem for de Sitter vacua
using this formalism.  In particular, restricting only to NSNS and
RR fluxes and O6/D6 sources we find that the critical quantity takes
the form
\begin{equation}
\frac{4 a c}{b^2} = (\mbox{const}) \sum_p \rho^{-p} \tilde{A}_p(M)\,
. \label{eq:CriticalNoGo}
\end{equation}
It is clear that the minimum of (\ref{eq:CriticalNoGo}) in the
$\rho$-direction is a runaway, $\rho\rightarrow \infty$, with $4 a
c/b^2 \rightarrow 0$.  Thus, de Sitter vacua cannot exist with these
ingredients.

In order to evade the no-go theorem of \cite{Hertzberg:2007wc} we
need to introduce different energy sources with different functional
dependence on the moduli $(\rho,\tau)$. Ideally, one would like to
not include {\it all} possible additional sources - it would be
helpful to know what is the minimal set of additional ingredients
needed in order to allow for de Sitter vacua.

Let us allow the possibility of non-zero, negative curvature of the
internal space in the scalar
potential. Indeed, we find now that the no-go theorem of
\cite{Hertzberg:2007wc} does not apply.  More precisely, the
critical quantity (\ref{eq:EvaCritical}) becomes,
\begin{equation}
\frac{4ac}{b^2} = (\mbox{const}) \sum_p \tilde{A_p} [ \tilde{C}_f
\rho^{2-p} + \tilde{A}_{H3}\rho^{-p}]\, .
\label{eq:CriticalEvadeNoGo}
\end{equation}
For simplicity, let us consider the coefficients $\tilde A_p, \tilde
C_f,\tilde A_{H3}$ to be pure constants, independent of any other
moduli. We now see that (\ref{eq:CriticalEvadeNoGo}) no longer has a
runaway potential for $\rho$ {\it as long as $\tilde{A}_p \neq 0$
for $p < 2$}, so including negative internal curvature and non-zero
$p<2$ flux are the minimal additional ingredients needed for de
Sitter vacua.  In IIA the latter statement translates into a
requirement that the IIA Romans mass parameter is non-zero (in IIB
this requirement suggests that RR $F_1$ flux is a necessary
ingredient), so we have a minimal set of requirements for de Sitter
vacua in IIA: {\bf In order to build de Sitter vacua at tree-level
in IIA, in addition to the usual RR and NSNS fluxes and O6/D6
sources, one must {\it minimally} have negative curvature spaces
\underline{and} non-zero Romans parameter.}

\begin{figure}[t]
\begin{center}
\includegraphics[scale=.45]{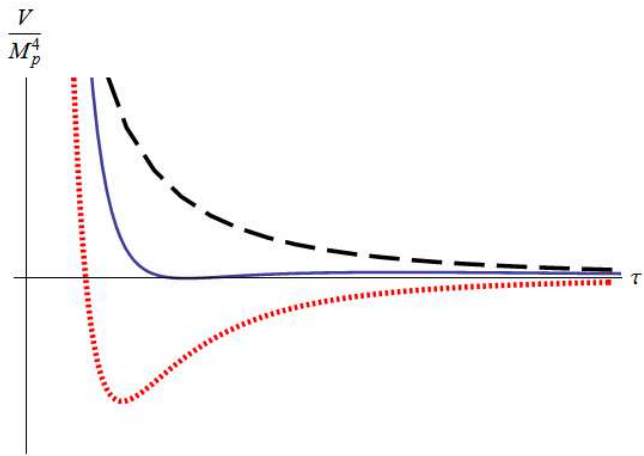}
\caption{\small Simple stable de Sitter solutions (solid line)
can be constructed by uplifting a stable AdS minimum
(dotted line) by negative curvature (dashed line).}
\label{fig:VUplifting}
\end{center}
\end{figure}

A simple intuitive way to understand this result is to investigate
the behavior of the potential (\ref{eq:potentialterms}) as a function
of $\tau$, as shown in Figure \ref{fig:VUplifting}.  Without the
negative curvature, the potential has an AdS minimum.  Adding in the
negative curvature acts as an uplifting term (slightly shifting the
minimum of the potential), lifting the AdS minimum to a dS one.
Clearly, we see a limitation on the amount of negative curvature
we can turn on in these models and still obtain stable de Sitter
vacua - for too large of the curvature, the de Sitter minimum of
Figure \ref{fig:VUplifting} disappears and the potential has
a runaway to a Minkowski vacuum at $\tau \rightarrow \infty$.
Notice that the argument given above did not require that the
solution is in the large volume and weak coupling regime - that
is an additional constraint that must be imposed upon candidate
solutions, and depends on the details of the construction.

Below we will discuss a simple model based on the compact
3-hyperboloid, which has constant negative curvature. Since the
compact 3-hyperboloid is rigid, it only has an overall scale modulus
$\rho$; we will see that indeed stable dS vacua can be found for
this model by minimizing the quantity (\ref{eq:CriticalEvadeNoGo}).

\section{de Sitter Vacua for Hyperbolic Space}

We will be considering massive IIA supergravity with sources. While
it is not yet clear how massive IIA supergravity emerges from a
perturbative string theory description (see \cite{Banks} for some
discussion), we will assume that such a supergravity limit exists.
For some interesting recent work on a string/M theory description of
the Roman's parameter see \cite{Gaiotto:2009mv}.

We will start with the action (\ref{eq:IIAAction}) with the string frame metric
\begin{eqnarray}
\d s_{10}^2 &=& g_{\mu\nu}^{(s)} \d x^\mu \d x^\nu + g_{mn} \d y^m \d y^n \\
&=& g_{\mu\nu}^{(s)} \d x^\mu \d x^\nu + \alpha' \rho\ d\tilde{s}_6^2
\label{eq:StringMetric}
\end{eqnarray}
where we factored out the overall breathing mode $\rho$ of the internal space,
\begin{equation}
\alpha'\rho \equiv \left[\frac{\int \sqrt{g_6}}{\int \sqrt{\tilde{g}_6}}\right]^{1/3}
\label{eq:rhodef}
\end{equation}
and we take the internal space to be the product of two identical compact
maximally symmetric 3-hyperboloids,
\begin{equation}
\d \tilde{s}_6^2 = \d{\mathbb H}_3^2(\Lambda) + \d\tilde{{\mathbb
H}}_3^2(\Lambda)\, .
\end{equation}
The 3-hyperboloid with curvature ${\mathcal R} = -\Lambda$ has the
metric
\begin{equation}
\d{\mathbb H}_3^2(\Lambda) = \tfrac{6}{\Lambda} \left(\d\varphi^2 +
\sinh^2(\varphi) \d\Omega_2^2\right)\, ,
\end{equation}
where $\varphi$ runs over the real line.  Algebraically, this space
is the coset $SO(3,1)/SO(3)$.  We will compactify the space by using
discrete $SO(3,1)$ identifications. By rescaling $\rho$ we can
always take $\Lambda=1$ \footnote{The overall curvature of the
internal space in units of $\alpha'$ equals $-2/\rho$.}.

The dimensionless volume of the internal space associated with
$d\tilde{s}_6^2$ is a constant, determined by the details
of the compactification of the space,
\begin{equation}
\tilde{V}_6=\int_{{(\mathbb H}_3\times{\mathbb H}_3)/\mathbb{Z}_2} \epsilon_3\wedge\tilde{\epsilon}_3
= \frac{e^{2\alpha}}{2} \,,
\end{equation}
where we defined the volume elements on the 3-hyperboloids as
$\epsilon_3,\tilde{\epsilon}_3$, respectively, and we denoted the
volume of the individual 3-hyperboloids as $e^{\alpha} \geq 1$. The
number $\alpha$ is discrete and bounded from below. It is related to
the specific discrete identifications one can make
\cite{Kaloper:2000jb}. In the integration domain we inserted a
$\mathbb{Z}_2$ from the O6-plane involution.  Similarly, we will
define the total physical volume of the internal space as
\begin{equation}
V_6 \equiv \int \sqrt{g_6}\, .
\end{equation}
Clearly, from (\ref{eq:rhodef}) the modulus $\rho$ is related to
these volumes as
\begin{equation}
\alpha'\rho = \left(V_6/\tilde{V}_6\right)^{1/3}\, .
\end{equation}
Finally, we will define the dimensionless field $\tau$ as,
\begin{equation}
\tau \equiv e^{-\phi} \rho^{3/2}\, ,
\end{equation}
where $e^{-\phi} = g_s^{-1}$ is the $10$-d string coupling
which appears in (\ref{eq:IIAAction}).

The dimensional reduction of the 10-dimensional Ricci scalar with
the Ansatz (\ref{eq:StringMetric}) leads to the non-canonical form
for the 4-d Ricci scalar,
\begin{equation}
\int \d^4x \sqrt{g_4^{(s)}} \left(\frac{\tau^2
\alpha'^3 \tilde{V}_6}{2\kappa_{10}^2}\right) \mathcal{R}_4^{(s)} + ...
\end{equation}
where $...$ includes terms depending on the curvature of the
internal space.
In order to bring the
4-dimensional curvature term into canonical Einstein-Hilbert form
$S_{EH} = \int \sqrt{|g|} 1/2\,M_{pl}^2 \mathcal{R}_4$ we will make
a conformal transformation $g_{\mu\nu}^{(s)} = (\tau/\tau_0)^{-2}
g_{\mu\nu}^{(E)}$, where $\tau_0$ is the stabilized value of $\tau$
in the vacuum.
Altogether, this gives a $10$-d metric in $4$-d Einstein frame and
canonical $4$-d Planck mass as,
\begin{eqnarray}
ds_{10}^2 &=& (\tau/\tau_0)^{-2} g_{\mu\nu}^{(E)} dx^\mu dx^\nu + \rho \alpha' d\tilde{s}_6^2 \\
M_p^2 &=& \frac{\tilde{V}_6 \alpha'^3 \tau_0^2}{\kappa_{10}^2} = \frac{V_{6,0}}{\kappa_{10}^2 g_{s,0}^2}\, ,
\end{eqnarray}
with $V_{6,0}, g_{s,0}$ the stabilized values of the total internal
volume and 10-d string coupling, respectively.

We further consider an O-plane spacetime action $R$ that mirrors one
3-hyperboloid to the other:
\begin{equation}
(z_1,z_2,z_3,\tilde{z}_1,\tilde{z}_2,\tilde{z}_3)
\longleftrightarrow
(\tilde{z}_1,\tilde{z}_2,\tilde{z}_3,z_1,z_2,z_3)\,.
\end{equation}
This corresponds to an O6-plane that is 4D space filling and wraps
the three-cycle in the internal manifold formed by the submanifold
$\Sigma_3^S$ invariant under the orientifold action:
\begin{equation}
\Sigma^S_3=(z_1,z_2,z_3,z_1,z_2,z_3)\,.
\end{equation}

\subsection{Scalar Potential}

Dimensionally reducing (\ref{eq:IIAAction}) on the above space leads
to a scalar potential for the moduli. Since the space is maximally
symmetric there are no deformations of the space aside from the
overall volume modulus $\rho$ \footnote{The scalars that deform the
shape are very massive and naturally stabilised at the value for
which the hyperboloid possesses its maximum symmetry.}. Further,
there are no non-trivial 2- or 4-cycles, so we cannot turn on $F_2$
or $F_4$ flux in the internal space, nor can we have flux moduli
descending from the the gauge potentials. Thus, we see that {\it
there are only two moduli in this model, $\rho$ and $\tau$}. As we
will see, it is due to the simplicity of this background that we
will easily find stable de Sitter solutions.

In our conventions the scalar potential $V$ is defined as the object
in the 4D action that appears as follows
\begin{equation}\label{potential}
S=\int \d
x_4\sqrt{g_4}\Bigl(\frac{M_p^2}{2}R_4-\frac{M_p^2}{2}G_{ij}\partial\phi^i\partial\phi^j
- V(\phi) \Bigr)\,,
\end{equation}
where we took the scalars to be dimensionless.

The potential energy contribution from the curvature is obtained
by simply reducing the internal curvature part of the 10-d Ricci
scalar, and going to 4-d Einstein frame,
\begin{equation}
V_{CURV}=\frac{M_p^2\tau_0^{2}}{\alpha'}\tau^{-2}\rho^{-1}\,.
\end{equation}

In massive IIA, the 0-form flux contributes a potential energy,
\begin{equation}
V_{F_0}=\frac{M_p^2\tau_0^{2}}{\alpha'}\tau^{-4}\rho^{3}\,\frac{f_0^2}{16\pi^2}\,.
\end{equation}
where we made use of the quantization of the 0-form flux
\begin{equation}
m_0 = \frac{f_0}{2\sqrt{2}\pi \sqrt{\alpha'}},\ \ f_0 \in {\mathbb Z}\,.
\end{equation}
We will also include RR 6-form flux,
\begin{equation}
F_6 = 2\,k_6\, \epsilon_3 \wedge \tilde \epsilon_3\, ,
\end{equation}
Using the quantization rule
\begin{equation}
\int_\frac{\mathbb H \times \mathbb H}{\mathbb Z_2} F_6= \frac{1}{\sqrt{2}} (2 \pi \sqrt{\alpha'})^5 f_6 , \ \ f_6 \in {\mathbb Z}\,.
\end{equation}
We find that
\begin{equation}
k_6=\frac{(2\pi \sqrt{\alpha'})^5}{\sqrt{2} e^{2\alpha}} f_6
\end{equation}
which contributes a potential energy
\begin{equation}
V_{F_6}=\frac{M_p^2 \tau_0^{2}}{\alpha'}\tau^{-4}\rho^{-3}\,\frac{(2\pi)^{10}\,f_6^2}{e^{4\alpha}}\,.
\end{equation}

Due to the orientifold projection, NSNS 3-form flux must thread  a
3-cycle which is odd under the ${\mathbb Z}_2$ involution. We write
\begin{equation}
H_3 = p \epsilon_3^A
\end{equation}
where $\epsilon_3^A$ is the corresponding form of the antisymmetric
3-cycle $\epsilon_3^A =\sqrt{2}(\epsilon_3 -\tilde{\epsilon}_3)$.
One can show that the volume of this cycle is
$\mbox{Vol}(\Sigma_3^A) =
\tfrac{1}{2}\sqrt{8}e^{\alpha}\alpha'^{3/2}$. The field strength is
related to the flux quantum number by
\begin{equation}
p = \frac{8\pi^2}{\sqrt{8}e^{\alpha}\sqrt{\alpha'}} h,\,\, h\in
{\mathbb Z}\,.
\end{equation}
From this we find the contribution to the energy to be
\begin{equation}
V_{H_3}=\frac{M_p^2 \tau_0^{2}}{\alpha'}\frac{8\pi^4
h^2}{e^{2\alpha}}\tau^{-2}\rho^{-3}h^2\,.
\end{equation}

To understand the effect of the O6 plane on the energy we recall
that the BPS O6 source term in the IIA action is (string
frame)\cite{DeWolfe:2005uu}
\begin{equation}\label{eq:O6action}
2(2\pi)^{-6}l_s^{-7}\int_{O6}
\e^{-\Phi}\sqrt{|g|}-2\sqrt{2}(2\pi)^{-6}l_s^{-7}\int_{O6} C_7\,,
\end{equation}
The O-plane contributes to the potential energy via the first term
in equation (\ref{eq:O6action}). We find
\begin{equation}
V_{O6}=-\frac{M_p^2\tau_0^{2}}{\alpha'}\e^{-\alpha}\,4\sqrt{8}\,\pi\,\tau^{-3}\,.
\end{equation}
Where we made use of the fact that the $\mathbb{Z}_2$-symmetric
3-cycle $\Sigma_3^S$ wrapped by the O6 plane has the volume,
$\mbox{Vol}(\Sigma_3^S)=\sqrt{8}e^{\alpha}l_S^3$.

The O6 plane also introduces a charge for $C_7$ through the second
term in in equation (\ref{eq:O6action}). This affects the Bianchi
identity for $F_2$
\begin{eqnarray}
\d F_2 &=& m_0 H_3 + 2\pi\sqrt{2}\,l_s\,\Sigma_3^{A}\,, \\
\d F_4 &=& -F_2\wedge H_3\,,
\end{eqnarray}
where $\Sigma_3^{A}\sim\epsilon^A$ is the antisymmetric 3-form
``orthogonal'' to the cycle $\Sigma_3^S$ that is wrapped by the
O6-plane ($\Sigma_3^{A}=\star_6\Sigma^S_3$). We also presented the
Bianchi identity for $F_4$. The associated ``tadpole relations'' are
found by integrating over a cycle and using Gauss' law. This gives
\begin{eqnarray}
\label{eq:H3Tadpole}
&& \int_{\Sigma_3^i}m_0 H_3 = -2\pi\sqrt{2}\,l_s\, \int_{\Sigma_3^i} \Sigma_3^{A}\,, \\
&& \int_{\Sigma_5}F_2\wedge H_3 = 0\,. \label{eq:F2Tadpole}
\end{eqnarray}%
The tadpole relation for $H_3$ flux (\ref{eq:H3Tadpole}) becomes
\begin{equation}
f_0 h = 2\,, \label{eq:H3TadpoleSimple}
\end{equation}
The other tadpole relations are satisfied trivially.

The potential energy from the NSNS flux, in terms of the Romans flux
parameter $f_0$, is
\begin{equation}
V_{H_3}=\frac{M_p^2\tau_0^{2}}{2} \frac{32\pi^4}{e^{2\alpha}f_0^2}\tau^{-2}\rho^{-3}\,.
\end{equation}
where now $f_0$ can only be $1$ or $2$ in order to satisfy the tadpole
relation (\ref{eq:H3TadpoleSimple}).

\subsection{Searching For De Sitter Vacua}

Collecting terms as in (\ref{eq:potentialterms}), and factoring out the overall
factor of $M_p^2\tau_0^{2}/\alpha'$, we have,
\begin{eqnarray}
\frac{\alpha'}{M_p^2\tau_0^{2}}a(\rho) &=& \frac{1}{\rho} + \frac{32\pi^4}{\e^{2\alpha}f_0^2}\rho^{-3}\,,\nonumber\\
\frac{\alpha'}{M_p^2\tau_0^{2}}b(\rho) &=& \e^{-\alpha}\,4\sqrt{8}\,\pi\,, \\
\frac{\alpha'}{M_p^2\tau_0^{2}} c(\rho) &=& \,\frac{f_0^2}{16\pi^2}\,\rho^{3}+
\,\frac{(2\pi)^{10}\,f_6^2}{e^{4\alpha}}
\,\rho^{-3}\nonumber\,.
\end{eqnarray}
The scalar potential is thus explicitly calculable in terms of the
microphysical parameters.  As discussed earlier, to find de Sitter
vacua we need only to find minima of the ``a,b,c" quantity near
unity,
\begin{equation}
\frac{4ac}{b^2}|_{minimum} \approx 1 + \delta
\end{equation}
for $\delta \ll 1$.
Using properties of these vacua, we have that
\begin{equation}
\tau=\frac{b}{2a}+\mathcal{O}(\delta)\,,
\end{equation}
from which we find that the stabilized value of the string
coupling is related to the overall volume,
\begin{equation}
g_s=\frac{\e^{\alpha}}{4\sqrt{2}\pi}\sqrt{\rho_0}
+\frac{4\sqrt{2}\pi^3}{\e^{\alpha}f_0^2}\frac{1}{(\sqrt{\rho_0})^3}\,.
\end{equation}
There is not much freedom in tuning various quantities and, as a
consequence, there is generically a trade off between having $g_s$
small and having a separation of scales between the Kaluza-Klein (KK) masses
and the moduli masses.
We shall
therefore examine a few solutions and
focus on i) the value of the string coupling, which determines
whether string loop corrections can be consistently ignored, ii) the
value of the internal volume  which determines whether
$\alpha'$-corrections can be consistently ignored and iii) the
masses of the moduli and the KK particles.

Before we proceed, let us elaborate further on point iii). In
performing our dimensional reduction to four dimensions, we
truncated the KK tower of states from the internal space, keeping
only the zero modes (which we have been calling ``moduli"). This
procedure is consistent if there exists a hierarchy of mass scales
between the KK modes and the zero modes. If no hierarchy exists then
the KK modes can contribute to the dynamics of the low energy theory
and the simple dimensional reduction to the zero modes does not give
a complete 4D effective theory \footnote{For example, in IIB flux
compactifications on Calabi-Yau spaces the backreaction of the
fluxes generate warping factors in the $10$-dimensional metric, and
in regions of strong warping there is generically no separation of
scales between the moduli and the KK modes so one cannot
consistently truncate the effective theory to the zero modes.
Further, the warping ends up modifying the dimensional reduction
procedure and affects the low energy effective theory and cannot be
ignored \cite{Giddings:2005ff, Shiu:2008ry, Frey:2008xw}.}.  In our
specific case, we expect setting the KK modes to zero is a \emph{consistent
truncation} in the sense of supergravity reduction. This is similar
to the Freund--Rubin vacua. The vacuum solution exists from a 10D
point of view but the KK modes are as important for the 4D physics
as the moduli. In other words, we expect the 10D equations of motion are
solved even though there is no clear separation of scales.
We expect to return to a more complete analysis of the 10D equations
of motion in future work.

Let us now look for an explicit solution. This can easily be done
numerically for various choices of the parameters; one set of
parameters which leads to a de Sitter vacua with small vacuum
energy:
\begin{table}[h]
\begin{center}
\begin{tabular}{l l l}
$f_0=2$, & $\frac{4ac}{b^2} \approx 1.03$, & $\frac{V_{dS}}{M_p^4} \approx 7.9\times 10^{-5}$\\
$f_6 = 8$, & $\rho_{dS} \approx 90.614$, & $\tau_{dS}\approx 1.47\times 10^3$\\
$\alpha \approx 0$, & &
\end{tabular}
\end{center}
\end{table}

\begin{figure}
\begin{center}
\includegraphics[scale=.45]{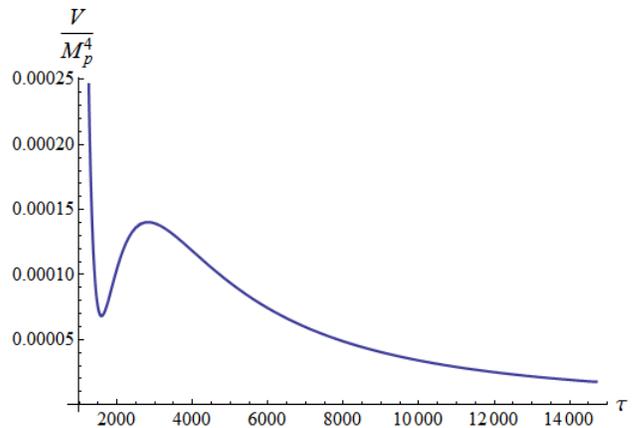}
\caption{\small Stable de Sitter vacua with small cosmological
constant can be obtained with minimal tree-level ingredients when
compactified on spaces of negative curvature.}
\label{fig:VdSStability}
\end{center}
\end{figure}

The potential for these flux choices is shown in Figure
\ref{fig:VdSStability}, which clearly illustrates a metastable de
Sitter vacuum in the $\tau$-direction as discussed earlier. In units
of $\alpha'$, the Planck mass is thus,
\begin{equation}
M_p^2 \approx \frac{5.59}{\alpha'}\, .
\end{equation}

Thus, we see that a main limitation of using a rigid compact
3-hyperboloid is that there are no parameterically small or large
numbers with which to simultaneously make the volume large and the
string coupling small.  Instead, one must rely on the precise
numerical factors in order to satisfy the consistency constraints.
For the parameters given above, we find
\begin{equation}
V_6 = \frac{1}{2} e^{2\alpha} \rho_{dS}^3 \approx 3.72\times 10^5,\,\,\, g_s \approx 0.56
\end{equation}
so we find that our solution is marginally within the large volume
and weak coupling regime. The actual string loop expansion parameter
for Type IIA orientifold in the present T-dual frame is $g_s$ times
a numerical factor which can be deduced (by T-duality) from that of
Type I string theory  discussed in
\cite{Caceres:1996is,Hebecker:2004ce}. Given the discrepancy in the
precise numerical factor in the literature, we take the more
conservative estimate  \footnote{We thank A.~Hebecker and
M.~Trapletti for a correspondence on this point.} about the region
of validity of perturbation theory in \cite{Hebecker:2004ce} which
suggests that the loop expansion parameter $\lambda$ is $\lambda
\leq \frac{g_s}{2 \sqrt{2}}$. Thus, our solution is in the
perturbative regime.

In order to compute the moduli masses, we first need to identify the
appropriate canonically normalized scalar fields.  It is
straightforward to see that the fields $\rho,\tau$ have kinetic
terms \cite{Hertzberg:2007wc},
\begin{equation}
S_{kinetic} = \int \sqrt{g_4^{(E)}} \left[M_p^2 \frac{(\partial \tau)^2}{\tau^2} + \frac{3 M_p^2}{4} \frac{(\partial \rho)^2}{\rho^2}\right]
\end{equation}
so that the canonically normalized scalar fields are
\begin{eqnarray}
\hat \tau &=& \sqrt{2} M_p \ln \tau \\
\hat \rho &=& \sqrt{\frac{3}{2}} M_p \ln \rho\, .
\end{eqnarray}
The square root of eigenvalues of the mass matrix computed as $M_{ij} =\frac{1}{2} \partial_i \partial_j V$,
for $i,j = \{\hat\tau,\hat\rho\}$, are, in units of $\alpha' = 1$,
\begin{equation}
\approx\{  0.22,\, 0.076 \}\, .
\end{equation}

The KK masses can be estimated in two ways; first, one can compute
the longest length $L$ around the 3-hyperboloid
\cite{Kaloper:2000jb}, and identify the KK mass as (again in units
of $\alpha'$)
\begin{equation}
m_{KK} \sim \frac{c_n}{\rho_{dS}^{1/2} L} \sim 0.17\, ,
\end{equation}
where we found $L \sim 0.61$ in string units and $c_n$ is some ${\mathcal O}(1)$ number.
Alternatively, one can estimate the KK mass scale from the overall volume,
\begin{equation}
m_{KK} \sim \frac{a_n}{V_6^{1/6}} \sim \frac{a_n}{(\alpha' \rho_{dS})^{1/2} \tilde{V}_6^{1/6}} \sim 0.11 \, .
\end{equation}
Clearly the two estimates are not too far off.

Comparing these KK masses to the moduli masses, we see that there is
no separation of scales between the KK masses and the moduli masses, as is expected
since our model does not have many tunable parameters with which to
create a separation of scales.
This feature is an artifact of the simple example we have chosen for illustration,
as the rigidity of the hyperbolic spaces also implies that there are fewer
adjustable parameters (like fluxes over a variety of cycles) to
separate these scales. It would be interesting to construct such examples from
 compactifications of other negatively Ricci curved spaces.

\section{De Sitter Vacua for Twisted 3-Tori}
\label{sec:Generalization}

In the previous section we examined a very simple background which
illustrates the minimal ingredients needed to construct stable,
three-level de Sitter solutions as seen in
(\ref{eq:CriticalEvadeNoGo}). A key aspect of this construction is
that the coefficients in (\ref{eq:CriticalEvadeNoGo}) were
constants, independent of the moduli. Unfortunately, models where
these coefficients are moduli-independent are not generic and may be
marginally within the large volume, weak coupling regime at best.

We can extend our analysis to include a simple set of models in
which the curvature of the internal space comes from a metric twist;
some of these geometries can be viewed as T-dual to spaces with NSNS
flux. In fact, for metric twists which are a product of two twisted
3-tori $G_3\times G_3$, all 3-dimensional Lie algebras were
classified by Bianchi so it is possible to exhaust all possibilities
(including the Nil-manifold considered in
\cite{Silverstein:2007ac}).

Let us now briefly consider the classification of twisted tori of
the form $G_3\times G_3$.

Given a parametrization of a Lie group $G$ we can define the
Maurer-Cartan forms via
\begin{equation}
g^{-1}\d g=\eta^aT_a\,,
\end{equation}
where the $T_a$ are the generators of the Lie algebra $\frak{G}$
associated to the Lie group $G$. Clearly $d (g^{-1}\d g)=-g^{-1}\d
g\wedge g^{-1}\d g$ and hence we can read of the Maurer-Cartan
equations
\begin{equation}
\d\eta^a=-f^a_{bc}\eta^b\wedge\eta^c\,,
\end{equation}
where $f^a_{bc}$ are the structure \emph{constants} of $\frak{G}$.
The metric on the Lie group is then defined via
\begin{equation}
\d s^2 = M_{ab}\eta^a\otimes \eta^b\,,
\end{equation}
where $M$ is any symmetric non-singular matrix. Since the $\eta^a$
are left invariant (under $g\rightarrow \Omega g$) this metric has a
left acting isometry group $G_L$. If $M$ coincides with the
Cartan--Killing metric we also have a right acting isometry such
that in total we have $G_L\times G_R$.

For a clear discussion on the classification of 3-dimensional Lie
algebras and the applications thereof in dimensional reduction of
supergravity theories we refer the reader to \cite{Roest:2004pk}.
The three-dimensional Lie algebras divide in two classes: class A
and class B according to following property of the Lie algebra
\begin{equation}
\text{class A}: f^n_{nm}=0\,,\qquad \text{class B}: f^n_{nm}\neq
0\,.
\end{equation}
It can be shown that reduction of the action on class B group
manifolds is inconsistent. Instead one has to reduce the equations
of motion and the result is that one obtains unusual theories with
the property that they do not allow a Lagrangian description, there
are only equations of motion \cite{Roest:2004pk}.

The class A Lie algebras are taken from \cite{Roest:2004pk} and
presented in table \ref{table:class A}.
\begin{table}
\begin{center}
\begin{tabular}{|c|c|c|}\hline
Bianchi type & Algebra & $(q_1,q_2,q_3)$  \\

\hline $I$     & $U(1)^3$ & $(0,0,0)$ \\

\hline $II$    & Heis$_3$ & $(0,0,Q_1)$    \\

\hline $VI_0$  & ISO(1,1) & $(0,-Q_1,Q_2)$    \\

\hline $VII_0$ & ISO(2)   & $(0,Q_1,Q_2)$    \\

\hline $VIII$  & SO(2,1)  & $(Q_1,-Q_2,Q_3)$    \\

\hline $IX$    & SO(3)    & $(Q_1,Q_2,Q_3)$    \\

\hline
\end{tabular}
\caption{\it The different class A 3D Lie algebras. The $Q_i$
represent the metric flux and are all positive numbers.}
\label{table:class A}
\end{center}
\end{table}
The $Q$'s denote the metric flux through the following relation
\begin{equation}
f_{bc}^a = \epsilon_{bcd} Q^{ad},\,\,\,\, Q =
\begin{pmatrix} q_1 &\ &\ \\ \ & q_2 &
\ \\ \ & \ & q_3 \end{pmatrix}\, .
\end{equation}

It is straightforward to perform the dimensional reduction and
search for stable de Sitter vacua for each of these twisted 3-tori.
Recall that in order to uplift to de Sitter solutions the
contribution to the potential energy from the curvature of the
internal manifold must be positive, which implies that the
stabilized value of the curvature of the manifold must be negative.
Starting from the bottom of the list in Table \ref{table:class A},
the SO$(2,1)$, SO$(3)$, and ISO$(2)$ groups do not allow stable de
Sitter vacua because the curvature for these spaces is not negative
definite, so the stabilized curvature cannot uplift the AdS vacuum.

The only possible candidate backgrounds, then, are Heis$_3$, studied
in \cite{Silverstein:2007ac}, and ISO$(1,1)$.  We have studied both
of these examples in detail, and have found that neither can support
a stable de Sitter solution for the minimal ingredients given in
(\ref{eq:CriticalEvadeNoGo}).  The reason is that the coefficients
in (\ref{eq:CriticalEvadeNoGo}) are now moduli dependent, so
the stabilization of $\rho$ depends on the stabilization of all of
the other moduli as well.  In particular, it is straightforward to
show that for the minimal ingredients above, there always exists
a runaway direction\footnote{We would especially like to thank Xi Dong for pointing
this out to us.}.

As a simple illustration, let us suppose that there is one
additional modulus $\phi$, and that the orientifold and $H_3$ flux
contributions are independent of this modulus (this will be the case
for the twisted 3-tori in which the orientifold maps the 3-manifolds
to each other); note also that the coefficient of the RR 0-form flux
$\tilde A_0$ is also moduli independent, since it does not involve
integrating over any cycles.  In general, then, the coefficients
have moduli dependence which we will parameterize as,
\begin{eqnarray}
\label{eq:GeometricScaling}
\tilde C_f &\sim & \phi^n \\
\tilde A_2 &\sim & \phi^{m_i} \\
\tilde A_4 &\sim & \phi^{2 \ell_i}\, .
\label{eq:RRScaling}
\end{eqnarray}
where we will assume $n > 0$ without loss of generality, and we have
allowed for the possibility of multiple different contributions of
$\phi$ to the RR forms, parameterized by different powers
$m_i,\ell_i$. The quantity $4ac/b^2$ becomes,
\begin{eqnarray}
\label{eq:quantity1mod}
\frac{4ac}{b^2} &\sim &  \mbox{const} + a_1 \phi^n \rho^2 + a_2 \phi^{n+m_i} \nonumber \\
    && + \frac{\left(a_3 \phi^{n+2\ell_i}+a_4 \phi^{m_i}\right)}{\rho^2} + {\mathcal O}(\rho^{-4},\rho^{-6})
\end{eqnarray}
\begin{eqnarray}
    &\sim & a_1 \tilde \phi^n + \frac{a_2 \tilde \phi^{n+m_i}+a_4 \tilde \phi^{m_i}}{\rho^{2 (1+m_i/n)}} + a_3 \frac{\tilde \phi^{n+2\ell_i}}{\rho^{4(1+\ell_i/n)}}
\label{eq:quantity1modScaled}
\end{eqnarray}
where in the third line we made the rescaling $\phi = \tilde
\phi/\rho^{2/n}$, and we ignored the constant and ${\mathcal
O}(\rho^{-4},\rho^{-6})$ contributions coming from the 6-form RR
flux because these tend to destabilize $\rho \rightarrow \infty$,
and so they will not be helpful in our stability analysis. That such
a rescaling is possible is due to the extra moduli dependence in
$\tilde C_f$; this field redefinition can remove the manifestly
positive power of $\rho$ coming from the product of the RR 0-form
and the geometric flux in (\ref{eq:quantity1mod}).  If it can be
shown that stable de Sitter vacua do not exist for the field
redefined quantity (\ref{eq:quantity1modScaled}) then vacua will not
exist for the original function (\ref{eq:quantity1mod}) either.

A positive power of $\rho$ can be regenerated in the other RR flux
terms by the field redefinition if their moduli dependence is just
right.
In particular, note that if $n+m_i \geq 0, n+\ell_i \geq 0$ for all
$m_i,\ell_i$, then after the field redefinition we only have
negative powers of $\rho$, and we
have a runaway $\rho \rightarrow \infty$ in trying to minimize
(\ref{eq:quantity1modScaled}). Indeed, it is clear that in order for
a positive power of $\rho$ to be regenerated in the expression
(\ref{eq:quantity1modScaled}) we need at least one of the moduli
dependences of the RR fluxes to satisfy $n + m_j < 0$ for some $m_j$
(alternatively, the same condition applies to the $\ell_j$).

For moduli coming from the metric of twisted 3-tori $M\sim \phi$, we generically expect contributions
of the order
\begin{eqnarray}
\tilde C_f &\sim & M^2 \sim \phi^2 \Rightarrow n = 2 \nonumber \\
\tilde A_2 &\sim & M^{-2} \sim \phi^{-2} \Rightarrow m_i = -2 \nonumber \\
\tilde A_4 &\sim & M^{-4} \sim \phi^{-4} \Rightarrow \ell_i = -2\, . \nonumber
\end{eqnarray}
These scalings are the dominant scalings - for example we generally
have $m_i,\ell_i \geq -2$, but we will not have stronger dependence
on $\phi$ than that listed above (e.g. we will not have $A_2,A_4$
scale with higher negative powers of $\phi$ than that listed above).
Also, note that these are the same scalings that are expected from
fractional Wilson lines as well, as they contribute to $F_2,F_4$
field strengths in similar ways as normal fluxes.  Thus, we find
that quite generally we expect $n + m_i, n+ \ell_i \geq 0$ for
twisted 3-tori - which suggests that with just metric flux,
O6-planes, RR and NSNS flux we do not have the sufficient
ingredients to find stable de Sitter vacua for twisted 3-tori.
Notice that while these ingredients were sufficient to find stable
de Sitter for the case when there were no additional moduli, when
additional moduli are present we find that there generically exists
a runaway direction $\rho \rightarrow \infty$.

As a more explicit check, let us consider the twisted 3-torus with
ISO$(1,1)$ metric twist.  There are two metric moduli, $L_1,L_2$.
It turns  out that the modulus $L_2$ only appears in the geometric
flux; after minimizing with respect to $L_2$, we find that the
remaining moduli dependence on $\phi \equiv 1/L_1$ is,
\begin{eqnarray}
\tilde C_f &\sim & 1/L_1^2 \sim \phi^2 \Rightarrow n = 2 \nonumber \\
\tilde A_2 & \sim & 1/L_1^4 + L_1^2 \sim \phi^4+\phi^{-2} \Rightarrow m_1 = 4, m_2 = -2 \nonumber \\
\tilde A_4 &\sim & L_1^4 + L_1^{-2} \sim \phi^2 + \phi^{-4} \Rightarrow \ell_1 = -2, \ell_2 = 1\, .\nonumber
\end{eqnarray}
Thus we see that indeed $n+m_i \geq 0, n + \ell_i \geq 0$ for all
$m_i,\ell_i$ in this example, so there does not exist a stable de
Sitter solution for finite $\rho$: There  always exists a field
redefinition, discussed above, which removes the positive power of
$\rho$ such that it is manifest that no stable de Sitter vacua
exist.

From the argument above it seems difficult  to obtain the desired
power of $m_i,\ell_i$ in the fluxes in order to have a stable de
Sitter minimum.  Instead, one can modify the moduli dependence in
the ``uplifting" energy by finding sources which have a different
$n$ dependence on $\phi$.  In particular, let us take a KK5-brane,
which has the same $\rho$-dependence as the geometric flux.  The
KK5-brane is wrapped on a 2-cycle in the internal space, thus it
will pick up moduli dependence of the form,
\begin{equation}
\tilde C_{KK5} \sim \int_{\Sigma_{2,KK5}} \sqrt{g_{ind}(\Sigma_{2,KK5})} \sim M \sim \phi \Rightarrow n' = 1\, .
\end{equation}
Combining the KK5-brane into the constructions above with geometric,
NSNS, and RR flux, we see that we can now easily satisfy the
requirement $n' + m_i < 0, n'+\ell_i < 0$ for the generic form of
$m_i,\ell_i$ that we expect. Thus, {\it negative curvature (e.g.
geometric flux), NSNS and RR fluxes (including 0-form), O-planes,
\underline{and} KK5-branes appear to be necessary ingredients in
order to find stable tree-level de Sitter vacua in IIA for twisted
3-tori.} Unfortunately, it is not clear if constructions with
KK5-branes in massive IIA are under complete control, particularly
in twisted tori backgrounds, since backreaction can be severe.

\section{Discussion}
It is a difficult problem to construct reliable stable de Sitter
vacua in string theory. We have argued that the minimal ingredients
needed to get de Sitter vacua in type IIA string theory are non-zero
Roman's parameter, RR, NSNS fluxes and negative internal curvature.
For the explicit example studied here in which the internal space is
a pair of compact, maximally symmetric 3-hyperboloids, we have shown
that one need only choose flux quanta appropriately to find de
Sitter vacua with small cosmological constant.  Because these
solutions have very few tunable parameters, however, we find that
our solution is marginally within the weak coupling regime, with
$g_s \approx 0.5$ \footnote{However, as pointed out earlier, we have
ignored subtleties with the definition of perturbative string theory
with non-zero Romans mass.}.

This simple model has just two moduli which are clearly
stabilised in a dS minimum. Furthermore we found that there is no
separation of scales in this simple model: the lightest KK modes are
of the same order of mass as the moduli. This is a possible drawback
if this solution should be considered as a semi-realistic vacuum.
But from the point of view of the dS/CFT correspondence
\cite{Strominger:2001pn} a separation of scales is not something
that is required, in the same way that the $AdS_5\times S^5$
solution has light KK modes.

We have also discussed the generalization of this simple model to
more general metric fluxes, by considering all twisted, orientable,
6-tori of the form $G_3\times G_3$, where $G_3$ represents the
covering space. There are 5 such families of twisted tori (not
including the normal torus): Heis$_3$, $\ISO(1,1)$, $\ISO(2)$,
$\SO(2,1)$ and $\SO(3)$. As discussed in Section
\ref{sec:Generalization}, only those twisted tori with negative
definite curvature provide the necessary uplifting energy to create
de Sitter solutions, immediately excluding ISO$(2)$, $\SO(2,1)$ and
$\SO(3)$. Constructions based on the Heis$_3$ background (also
sometimes called the Nil 3-manifold) and ISO$(1,1)$ background
appear to require additional ingredients such as KK5-branes in order
to stabilize all of the metric moduli, as in
\cite{Silverstein:2007ac}.  We showed why this is true by
considering a simple field redefinition which leads to a runaway
minimum for $\rho$ unless KK5-branes are included.

While we have shown that minimal stable de Sitter solutions do not
exist for manifolds which are a product of twisted 3-tori in IIA,
our analysis does not rule out the possibility that the minimal set
of ingredients can lead to stable de Sitter solutions for
6-manifolds with negative curvature which can not be decomposed as
the product of 3-tori. The benefit of simple dS solutions from
twisted tori is that susy is not broken at the KK scale such that
the dS solution is a spontaneous broken state in an $\mathcal{N}=4$
gauged supergravity coupled to six vectormultiplets
\cite{Andrianopoli:2003sa, Angelantonj:2003rq, Angelantonj:2003up,
Dall'Agata:2005ff}. The latter theories are rigid in the sense that
the number of vector multiplets and the gauge group almost fully
determine all interactions. The only freedom resides in the values
of the gauge coupling constants and so called ``de Roo--Wagemans
angles'' (\emph{aka} $\SU(1,1)$ angles) \cite{deRoo:1985jh}.
Especially those angles play a central role in providing de Sitter
vacua in extended supergravity \cite{deRoo:2002jf, deRoo:2003rm,
Fre:2002pd, deRoo:2006ms}.  This construction could offer a string
theory embedding of the `de Roo - Wagemans ' angles which were
introduced in supergravity but their relation with string theory is
so far only established in the examples of \cite{Derendinger:2004jn}
(and effectively obtained in an $\mathcal{N}=1$ context). Finally
having the gauged supergravity description at hand allows an easier
derivation of the explicit mass matrix, such that we can work out
whether the minimal simple de Sitter vacua are stable with respect
to all fluctuations. Embedding our constructions into 4D gauged
supergravity also enables us to compute the gravitino mass, and
check whether the bound in \cite{Kallosh:2004yh} applies.

It would also be interesting to study whether these simple de Sitter
vacua exist in type IIB backgrounds as well. A geometric background
can be looked for using a similar derivation as in section
\ref{scalarpotential} to derive the necessary ingredients to obtain
de Sitter vacua. From this it becomes apparent that the minimal
ingredients are NSNS and geometric flux, Op-plane sources and $F_1$
flux (not necessarily all together). We thus expect minimal simple
de Sitter solutions to exist in IIB as well, although additional
ingredients may be needed in models with more than one moduli, as
seen here. Nevertheless, this may give some insights into
stabilization mechanisms for K\"ahler moduli in type IIB which do
not rely on non-perturbative effects, as well as identify possible
uplifting sources of energy.

Furthermore, it could be fruitful to use minimal de Sitter solutions
as starting points to construct large-field inflation models
\cite{Silverstein:2008sg,McAllister:2008hb} and particle physics
constructions \cite{Camara:2005dc,Marchesano:2006ns} explored in
similar backgrounds. As the low energy chiral particle physics
spectrum depends only on the topological data of cycles on which the
D-branes are wrapped, the machineries developed for intersecting
D-brane models \cite{Blumenhagen:2005mu} can be readily adopted to
simple extensions of toroidal backgrounds. When embedded into a
single framework, these investigations may thus allow us to study
the interplay between cosmology and particle physics.

Finally, the simplicity of our background solution $dS_4\times
\mathbb{H}_3\times \mathbb{H}_3$ is of the same simplicity as the
Freund-Rubin vacua $AdS_n\times S^m$ which allow explicit tests of
the AdS/CFT conjecture. The background presented here is therefore a
good starting point for testing a hypothetical $dS/CFT$
correspondence.

We hope to return to all these exciting directions in the future.

\section*{Acknowledgments}
We would like to thank A.~Collinucci, M.~de Roo, A.~Frey, A.~Hebecker,
A.~Maloney, S.~Kachru, R.~Reid-Edwards, E.~Silverstein, G.~Starkman,
M.~Trapletti,
M.~Trigiante,
S.~Trivedi, and
D.~Westra for helpful discussions. We are especially grateful to X.~Dong
for pointing out an error in an earlier version of this work.
GS thanks the Stanford Institute
for Theoretical Physics and SLAC for hospitality and support while
this work was written. The work of GS and SSH is supported in part
by NSF CAREER Award No. PHY-0348093, DOE grant DE-FG-02-95ER40896, a
Research Innovation Award and a Cottrell Scholar Award from Research
Corporation, a Vilas Associate Award from the University of
Wisconsin, and a John Simon Guggenheim Memorial Foundation
Fellowship. TVR thanks the Junta de Andaluc\'ia for financial
support. BU is supported by NSERC and by fellowships from the
Institute for Particle Physics (Canada) and Lorne Trottier (McGill).

\appendix

\section{``a,b,c" de Sitter vacua}
\label{sec:appendix}

In this section we present a simple derivation,
originally discussed in \cite{Silverstein:2007ac},
that searching for de Sitter critical points with small vacuum energy
of the potential
\begin{equation}
V= a(\phi^i) g^2 - b(\phi^i) g^3+ c(\phi^i) g^4\,.
\label{eq:AppendixPotential}
\end{equation}
with respect to all of the moduli $(g,\phi^i)$ corresponds
to finding critical points of the quantity $4ac/b^2 \approx 1$ as a function
of the other moduli $\phi^i$.
Note that in this notation $g = 1/\tau$ corresponds to the potential used
above in (\ref{eq:potentialterms}).

Minimizing the potential (\ref{eq:AppendixPotential}) in the $g,\phi^i$ directions leads
to the equations,
\begin{eqnarray}
\label{eq:phicriticaleqn}
(\partial_{\phi_i}a)  - (\partial_{\phi_i}b) g_{0}+(\partial_{\phi_i}c)g^{2}_{0}&=& 0 \\
\left(g^2-\frac{3b}{4c} g +\frac{a}{2c}\right)\mid _ {g _{0},\phi^i_0}&=& 0
\label{eq:gcriticaleqn}
\end{eqnarray}
where $(g_0,\phi^i_0)$ are the putative stabilized values of the moduli, and
we will denote $a_0,b_0,c_0$ as the corresponding stabilized values of the functions
appearing in the potential (\ref{eq:AppendixPotential}).
The expression (\ref{eq:gcriticaleqn}) can be solved for $g_0$ in terms of the
stabilized values of $a,b,c$
\begin{equation}
g_0 = \frac {3b_0}{8 c_0}\pm \frac{b_0}{2} \sqrt{\frac{9 }{16 c_0^2} - \frac{1} {2 c_0^2} (\frac{4 a_0 c_0}{b_0^2})}\, .
\end{equation}
When $\frac {4 a_0 c_0}{b_0^2}\simeq 1+\delta$, with $\delta<<1$,
we have two possible solutions
\begin{eqnarray}
\label{eq:g0soln}
g_0 &\simeq & \frac{b_0}{2 c_0} (1+\frac{1}{8} \delta) \\
g_0 &\simeq & \frac{b_0}{4 c_0} (1-\frac{1}{4} \delta)\, .
\end{eqnarray}
The solution $g_0 \simeq b_0/(2 c_0)$
corresponds to a local minimum in the $g$ direction with positive vacuum energy proportional
to $\delta$ (and is thus the de Sitter vacua we are interested in)
\begin{equation}
V_{min}|_{g_0=\frac{b_0}{2c_0}(1+1/8\delta),\phi^i_0} = g_0 c_0 \delta
\end{equation}
so we will focus on this solution henceforth.

Now, searching for critical points of $4ac/b^2$ with respect to the moduli $\phi^i$, we obtain (after some
basic manipulation)
\begin{equation}
\partial_{\phi_i}a -(\frac{2a}{b}) \partial_{\phi_i}b + \frac {a}{c} \partial_{\phi_i} c = 0
\label{eq:abccriticalequation}
\end{equation}
The approach of finding critical points of $4ac/b^2 \approx 1$ is equivalent
to finding local de Sitter minima of the entire potential (\ref{eq:AppendixPotential})
if solving the expression (\ref{eq:abccriticalequation})
is identical to solving the expression (\ref{eq:phicriticaleqn}).  It is clear that this
is true only if
\begin{equation}
g_0 = \frac{2a_0}{b_0},\,\,\, g_0^2 = \frac{a_0}{c_0}\, .
\end{equation}
But using $4a_0c_0/b_0^2 = 1 + \delta$, it is easy to see in fact that
\begin{equation}
\frac{2a_0}{b_0} = \frac{b_0}{2 c_0}(1+\delta),\,\,\,\, \frac{a_0}{c_0} = \left(\frac{b_0}{2 c_0}\right)^2 (1+\delta)
\end{equation}
which both imply the solution (\ref{eq:g0soln}), $g_0 = b_0/(2 c_0) + {\mathcal O}(\delta)$.
Thus, the critical point equations (\ref{eq:abccriticalequation}) and (\ref{eq:phicriticaleqn})
are in fact identical when $4ac/b^2 \approx 1 + \delta$ when $\delta \ll 1$, so when searching for
de Sitter vacua with small vacuum energy it is sufficient to search for critical points of $4ac/b^2 \approx 1$.
The advantage of this approach is that in many cases (as discussed above) it is clear simply by inspection
when $4ac/b^2$ cannot be minimized at all for finite values of the moduli; thus, these cases can be
immediately ruled out as candidate de Sitter vacua
without needing to solve the entire system of equations (\ref{eq:phicriticaleqn},\ref{eq:gcriticaleqn})
for every specific example.

\bibliography{groups}

\providecommand{\href}[2]{#2}\begingroup\raggedright\begin{thebibliography}{10}

\bibitem{Perlmutter:1998np}
{\bf Supernova Cosmology Project} Collaboration, S.~Perlmutter {\em et al.},
  {\em {Measurements of Omega and Lambda from 42 High-Redshift Supernovae}},
  Astrophys. J. {\bf 517} (1999) 565--586
[\href{http://www.arXiv.org/abs/astro-ph/9812133}{astro-ph/9812133}].

\bibitem{Riess:1998cb}
{\bf Supernova Search Team} Collaboration, A.~G. Riess {\em et al.},  {\em
  {Observational Evidence from Supernovae for an Accelerating Universe and a
  Cosmological Constant}}, Astron. J. {\bf 116} (1998) 1009--1038
[\href{http://www.arXiv.org/abs/astro-ph/9805201}{astro-ph/9805201}].

\bibitem{Kachru:2003aw}
S.~Kachru, R.~Kallosh, A.~Linde and S.~P. Trivedi,  {\em {De Sitter vacua in
  string theory}}, Phys. Rev. {\bf D68} (2003) 046005
[\href{http://www.arXiv.org/abs/hep-th/0301240}{hep-th/0301240}].

\bibitem{Maloney:2002rr}
A.~Maloney, E.~Silverstein and A.~Strominger,  {\em {De Sitter space in
  noncritical string theory}},
[\href{http://www.arXiv.org/abs/hep-th/0205316}{hep-th/0205316}].

\bibitem{Balasubramanian:2005zx}
V.~Balasubramanian, P.~Berglund, J.~P. Conlon and F.~Quevedo,  {\em
  {Systematics of moduli stabilisation in Calabi-Yau flux compactifications}},
  JHEP {\bf 03} (2005) 007
[\href{http://www.arXiv.org/abs/hep-th/0502058}{hep-th/0502058}].

\bibitem{Saltman:2004sn}
A.~Saltman and E.~Silverstein,  {\em {The scaling of the no-scale potential and
  de Sitter model building}}, JHEP {\bf 11} (2004) 066
[\href{http://www.arXiv.org/abs/hep-th/0402135}{hep-th/0402135}].

\bibitem{Saltman:2004jh}
A.~Saltman and E.~Silverstein,  {\em {A new handle on de Sitter
  compactifications}}, JHEP {\bf 01} (2006) 139
[\href{http://www.arXiv.org/abs/hep-th/0411271}{hep-th/0411271}].

\bibitem{Saueressig:2005es}
F.~Saueressig, U.~Theis and S.~Vandoren,  {\em {On de Sitter vacua in type IIA
  orientifold compactifications}}, Phys. Lett. {\bf B633} (2006) 125--128
[\href{http://www.arXiv.org/abs/hep-th/0506181}{hep-th/0506181}].

\bibitem{Davidse:2005ef}
M.~Davidse, F.~Saueressig, U.~Theis and S.~Vandoren,  {\em {Membrane instantons
  and de Sitter vacua}}, JHEP {\bf 09} (2005) 065
[\href{http://www.arXiv.org/abs/hep-th/0506097}{hep-th/0506097}].

\bibitem{Silverstein:2007ac}
E.~Silverstein,  {\em {Simple de Sitter Solutions}}, Phys. Rev. {\bf D77}
  (2008) 106006
[\href{http://www.arXiv.org/abs/0712.1196}{0712.1196}].

\bibitem{Palti:2008mg}
E.~Palti, G.~Tasinato and J.~Ward,  {\em {WEAKLY-coupled IIA Flux
  Compactifications}}, JHEP {\bf 06} (2008) 084
[\href{http://www.arXiv.org/abs/0804.1248}{0804.1248}].

\bibitem{Misra:2007yu}
A.~Misra and P.~Shukla,  {\em {Area Codes, Large Volume (Non-)Perturbative
  alpha'- and Instanton - Corrected Non-supersymmetric (A)dS minimum, the
  Inverse Problem and Fake Superpotentials for Multiple-
  Singular-Loci-Two-Parameter Calabi-Yau's}}, Nucl. Phys. {\bf B799} (2008)
  165--198
[\href{http://www.arXiv.org/abs/0707.0105}{0707.0105}].

\bibitem{Susskind:2003kw}
L.~Susskind,  {\em {The anthropic landscape of string theory}},
\href{http://www.arXiv.org/abs/hep-th/0302219}{hep-th/0302219}.

\bibitem{Bousso:2000xa}
R.~Bousso and J.~Polchinski,  {\em {Quantization of four-form fluxes and
  dynamical neutralization of the cosmological constant}}, JHEP {\bf 06} (2000)
  006
[\href{http://www.arXiv.org/abs/hep-th/0004134}{hep-th/0004134}].

\bibitem{Feng:2000if}
J.~L. Feng, J.~March-Russell, S.~Sethi and F.~Wilczek,  {\em {Saltatory
  relaxation of the cosmological constant}}, Nucl. Phys. {\bf B602} (2001)
  307--328
[\href{http://www.arXiv.org/abs/hep-th/0005276}{hep-th/0005276}].

\bibitem{Weinberg:1987dv}
S.~Weinberg,  {\em {Anthropic Bound on the Cosmological Constant}}, Phys. Rev.
  Lett. {\bf 59} (1987)
2607.

\bibitem{Strominger:2001pn}
A.~Strominger,  {\em {The dS/CFT correspondence}}, JHEP {\bf 10} (2001) 034
[\href{http://www.arXiv.org/abs/hep-th/0106113}{hep-th/0106113}].

\bibitem{DeWolfe:2005uu}
O.~DeWolfe, A.~Giryavets, S.~Kachru and W.~Taylor,  {\em {Type IIA moduli
  stabilization}}, JHEP {\bf 07} (2005) 066
[\href{http://www.arXiv.org/abs/hep-th/0505160}{hep-th/0505160}].

\bibitem{Hertzberg:2007wc}
M.~P. Hertzberg, S.~Kachru, W.~Taylor and M.~Tegmark,  {\em {Inflationary
  Constraints on Type IIA String Theory}}, JHEP {\bf 12} (2007) 095
[\href{http://www.arXiv.org/abs/arXiv:0711.2512 [hep-th]}{arXiv:0711.2512
  [hep-th]}].

\bibitem{Burgess:2003ic}
C.~P. Burgess, R.~Kallosh and F.~Quevedo,  {\em {de Sitter string vacua from
  supersymmetric D-terms}}, JHEP {\bf 10} (2003) 056
[\href{http://www.arXiv.org/abs/hep-th/0309187}{hep-th/0309187}].

\bibitem{Villadoro:2005yq}
G.~Villadoro and F.~Zwirner,  {\em {de Sitter vacua via consistent D-terms}},
  Phys. Rev. Lett. {\bf 95} (2005) 231602
[\href{http://www.arXiv.org/abs/hep-th/0508167}{hep-th/0508167}].

\bibitem{Silverstein:2008sg}
E.~Silverstein and A.~Westphal,  {\em {Monodromy in the CMB: Gravity Waves and
  String Inflation}}, Phys. Rev. {\bf D78} (2008) 106003
[\href{http://www.arXiv.org/abs/0803.3085}{0803.3085}].

\bibitem{Covi:2008ea}
L.~Covi {\em et al.},  {\em {de Sitter vacua in no-scale supergravities and
  Calabi-Yau string models}}, JHEP {\bf 06} (2008) 057
[\href{http://www.arXiv.org/abs/0804.1073}{0804.1073}].

\bibitem{Banks}
T.~Banks and K.~van~den Broek,  {\em {Massive IIA flux compactifications and
  U-dualities}}, JHEP {\bf 03} (2007) 068
[\href{http://www.arXiv.org/abs/hep-th/0611185}{hep-th/0611185}].

\bibitem{Gaiotto:2009mv}
D.~Gaiotto and A.~Tomasiello,  {\em {The gauge dual of Romans mass}},
\href{http://www.arXiv.org/abs/0901.0969}{0901.0969}.

\bibitem{Kaloper:2000jb}
N.~Kaloper, J.~March-Russell, G.~D. Starkman and M.~Trodden,  {\em {Compact
  hyperbolic extra dimensions: Branes, Kaluza-Klein modes and cosmology}},
  Phys. Rev. Lett. {\bf 85} (2000) 928--931
[\href{http://www.arXiv.org/abs/hep-ph/0002001}{hep-ph/0002001}].

\bibitem{Giddings:2005ff}
S.~B. Giddings and A.~Maharana,  {\em {Dynamics of warped compactifications and
  the shape of the warped landscape}}, Phys. Rev. {\bf D73} (2006) 126003
[\href{http://www.arXiv.org/abs/hep-th/0507158}{hep-th/0507158}].

\bibitem{Shiu:2008ry}
G.~Shiu, G.~Torroba, B.~Underwood and M.~R. Douglas,  {\em {Dynamics of Warped
  Flux Compactifications}}, JHEP {\bf 06} (2008) 024
[\href{http://www.arXiv.org/abs/0803.3068}{0803.3068}].

\bibitem{Frey:2008xw}
A.~R. Frey, G.~Torroba, B.~Underwood and M.~R. Douglas,  {\em {The Universal
  Kaehler Modulus in Warped Compactifications}},
\href{http://www.arXiv.org/abs/0810.5768}{0810.5768}.

\bibitem{Caceres:1996is}
E.~Caceres, V.~S. Kaplunovsky and I.~M. Mandelberg,  {\em {Large-volume string
  compactifications, revisited}}, Nucl. Phys. {\bf B493} (1997) 73--100
[\href{http://www.arXiv.org/abs/hep-th/9606036}{hep-th/9606036}].

\bibitem{Hebecker:2004ce}
A.~Hebecker and M.~Trapletti,  {\em {Gauge unification in highly anisotropic
  string compactifications}}, Nucl. Phys. {\bf B713} (2005) 173--203
[\href{http://www.arXiv.org/abs/hep-th/0411131}{hep-th/0411131}].

\bibitem{Roest:2004pk}
D.~Roest,  {\em {M-theory and gauged supergravities}}, Fortsch. Phys. {\bf 53}
  (2005) 119--230
[\href{http://www.arXiv.org/abs/hep-th/0408175}{hep-th/0408175}].

\bibitem{Andrianopoli:2003sa}
L.~Andrianopoli, S.~Ferrara and M.~Trigiante,  {\em {Fluxes, supersymmetry
  breaking and gauged supergravity}},
[\href{http://www.arXiv.org/abs/hep-th/0307139}{hep-th/0307139}].

\bibitem{Angelantonj:2003rq}
C.~Angelantonj, S.~Ferrara and M.~Trigiante,  {\em {New D = 4 gauged
  supergravities from N = 4 orientifolds with fluxes}}, JHEP {\bf 10} (2003)
  015
[\href{http://www.arXiv.org/abs/hep-th/0306185}{hep-th/0306185}].

\bibitem{Angelantonj:2003up}
C.~Angelantonj, S.~Ferrara and M.~Trigiante,  {\em {Unusual gauged
  supergravities from type IIA and type IIB orientifolds}}, Phys. Lett. {\bf
  B582} (2004) 263--269
[\href{http://www.arXiv.org/abs/hep-th/0310136}{hep-th/0310136}].

\bibitem{Dall'Agata:2005ff}
G.~Dall'Agata and S.~Ferrara,  {\em {Gauged supergravity algebras from twisted
  tori compactifications with fluxes}}, Nucl. Phys. {\bf B717} (2005) 223--245
[\href{http://www.arXiv.org/abs/hep-th/0502066}{hep-th/0502066}].

\bibitem{deRoo:1985jh}
M.~de~Roo and P.~Wagemans,  {\em {Gauge matter coupling in N=4 supergravity}},
  Nucl. Phys. {\bf B262} (1985)
644.

\bibitem{deRoo:2002jf}
M.~de~Roo, D.~B. Westra and S.~Panda,  {\em {De Sitter solutions in N = 4
  matter coupled supergravity}}, JHEP {\bf 02} (2003) 003
[\href{http://www.arXiv.org/abs/hep-th/0212216}{hep-th/0212216}].

\bibitem{deRoo:2003rm}
M.~de~Roo, D.~B. Westra, S.~Panda and M.~Trigiante,  {\em {Potential and
  mass-matrix in gauged N = 4 supergravity}}, JHEP {\bf 11} (2003) 022
[\href{http://www.arXiv.org/abs/hep-th/0310187}{hep-th/0310187}].

\bibitem{Fre:2002pd}
P.~Fre, M.~Trigiante and A.~Van~Proeyen,  {\em {Stable de Sitter vacua from N =
  2 supergravity}}, Class. Quant. Grav. {\bf 19} (2002) 4167--4194
[\href{http://www.arXiv.org/abs/hep-th/0205119}{hep-th/0205119}].

\bibitem{deRoo:2006ms}
M.~de~Roo, D.~B. Westra and S.~Panda,  {\em {Gauging CSO groups in N = 4
  supergravity}}, JHEP {\bf 09} (2006) 011
[\href{http://www.arXiv.org/abs/hep-th/0606282}{hep-th/0606282}].

\bibitem{Derendinger:2004jn}
J.-P. Derendinger, C.~Kounnas, P.~M. Petropoulos and F.~Zwirner,  {\em
  {Superpotentials in IIA compactifications with general fluxes}}, Nucl. Phys.
  {\bf B715} (2005) 211--233
[\href{http://www.arXiv.org/abs/hep-th/0411276}{hep-th/0411276}].

\bibitem{Kallosh:2004yh}
R.~Kallosh and A.~Linde,  {\em {Landscape, the scale of SUSY breaking, and
  inflation}}, JHEP {\bf 12} (2004) 004
[\href{http://www.arXiv.org/abs/hep-th/0411011}{hep-th/0411011}].

\bibitem{McAllister:2008hb}
L.~McAllister, E.~Silverstein and A.~Westphal,  {\em {Gravity Waves and Linear
  Inflation from Axion Monodromy}},
\href{http://www.arXiv.org/abs/0808.0706}{0808.0706}.

\bibitem{Camara:2005dc}
P.~G. Camara, A.~Font and L.~E. Ibanez,  {\em {Fluxes, moduli fixing and
  MSSM-like vacua in a simple IIA orientifold}}, JHEP {\bf 09} (2005) 013
[\href{http://www.arXiv.org/abs/hep-th/0506066}{hep-th/0506066}].

\bibitem{Marchesano:2006ns}
F.~Marchesano,  {\em {D6-branes and torsion}}, JHEP {\bf 05} (2006) 019
[\href{http://www.arXiv.org/abs/hep-th/0603210}{hep-th/0603210}].

\bibitem{Blumenhagen:2005mu}
R.~Blumenhagen, M.~Cvetic, P.~Langacker and G.~Shiu,  {\em {Toward realistic
  intersecting D-brane models}}, Ann. Rev. Nucl. Part. Sci. {\bf 55} (2005)
  71--139
[\href{http://www.arXiv.org/abs/hep-th/0502005}{hep-th/0502005}].

\end{thebibliography}\endgroup
\bibliographystyle{utphysmodb}

\end{document}